# Hierarchical Structure Design and Primary Energy Dispatching Strategy of Grid Energy Router

Meifu Chen [a], Mingchao Xia [a,*], and Qifang Chen [a]

[a] *School of Electrical Engineering, Beijing Jiaotong University, Beijing 100044, China.*

*Abstract*—As a core device of energy Internet, the energy router is deployed to manage energy flow between the renewable energy and electric grid. In this paper, a hierarchical structure of grid energy router is proposed to greatly facilitate peer-to-peer energy sharing among energy routers. It can be placed at critical buses to make active distribution networks develop into multiple interconnected prosumer-based autonomous systems. To alleviate the mismatch between the medium-time dispatch and device-level control caused by the forecast error of distributed generation, a bi-level primary energy dispatching strategy is proposed to fully utilize the energy buffer of multiple grid energy routers. The power variation in short-time scale is well suppressed by sharing energy buffer in the upper-level control, and the energy buffer is further optimized to better absorb the variation. Combining measured information, the lower-level control is designed to track the optimized instruction of energy buffer in real-time scale, which is a distributed process. The power flow constraint is assumed to be handled by medium-time dispatch, and the current constraint of the device is only taken into consideration. Finally, simulation results demonstrate the effectiveness of proposed hierarchical structure and primary energy dispatching strategy of the grid energy router.

## 1. Introduction

As the core equipment of flexible energy management in the energy Internet (EI), an energy router (ER) can be utilized to achieve large penetration of distributed generations (DG) and alleviate the problems caused by disorder charging of electric vehicles (EVs) [1, 2, 3]. A solid-state transformer-based structure is typically adopted by the present ER, and it is employed as an integrated interface to replace traditional transformer [4]. Such a kind of ER can be viewed as a transformer energy router (TER), and can provide abundant plug-and-play interfaces for DG and load. The studies on TER mainly focus on voltage matching, modular design, fault isolation, control strategy design and performance optimization of the equipment [5, 6, 7].

Studies on peer-to-peer energy sharing among ERs from the aspect of structure of ER are far less. In [8], ER is viewed as the execution device of the agent to realize the distributed energy sharing strategy among microgrids. In [9], the concept of different applications of ERs in DN is put forward, excluding that for convenient peer-to-peer energy sharing. In [10], the ER is employed for flexible power flow control of the DN, but only the calculation model of power flow is fully taken into consideration. In [11, 12, 13], optimal energy routing among ERs based on line loss is derived assuming that peer-to-peer energy sharing can be flexibly achieved by ER. In [14], a flexible multi-state switch is deployed to achieve energy sharing among different feeders in DN, but the effective energy sharing is limited to the end of connected feeders, and the flexibility of active energy management is finite without energy buffer devices. In sum, peer-to-peer energy sharing with different needs cannot be conveniently achieved through the present structure. Besides, such issues as power flow congestion, inverse direction of power flow and overvoltage of certain buses have become vital constraints on consumption of DG in active DN (ADN) which is mostly based on radial structure [15, 16, 17]. Those serious issues can be greatly resolved by peer-to-peer energy sharing as well. Thus, peer-to-peer energy sharing of different ERs is of great significance to facilitate the maximum consumption of DG from the entire ADN and meet different requirements of consumers of electricity [18, 19].

For the energy management of ERs, research mainly focuses on energy balance in real-time scale or energy dispatch. For energy management in real-time scale, different operation modes are designed to meet the energy balance between connected systems and the electric grid [20, 21, 22, 23, 24]. As an effective approach to derive multiple targets at the same time in real-time scale, fuzzy logic control (FLC) has been deployed in the energy balance of ERs. In [22], a fuzzy logic control (FLC) based strategy is deployed to manage the state of charge (SOC) of the energy buffer device equipped with a hybrid energy storage system. In [23], the factor of price of electricity is included in the FLC-based energy management of the energy buffer device to achieve economic operation. In [24], FLC-based energy management is employed to generate the operation mode switch signal, taking the state of energy buffer and DG into consideration. However, much attention is paid to stable operation and power variation within a single ER in real-time scale, which may lead to difficulty in implementing the upper energy dispatch results. Actually, there exists relatively complementary characteristics of the sources [25]. The power variation of the entire ER-based ADN should be shared to alleviate the power variation within a single ER in device-level control strategy, and thus the energy dispatch instructions can be more consist with the reality, and the maximum consumption of DG can be achieved.

For the optimization dispatch strategy, a hierarchical strategy is proposed to optimize the operation and energy route of the entire ER-based network in the multi-carrier energy system [26]. In [27], the optimization dispatch of EI solved by deep reinforcement learning approach is proposed to improve the optimal scheme such as the utilization of the energy buffer device. In [15, 16, 17], energy sharing within ER-based ADN is achieved by distributed strategy. However, those optimization strategies essentially belong to a medium-time dispatch in a medium-time scale and are sensitive to forecast errors. In order to handle the power variation in a short-time scale, model predictive control (MPC) is usually employed in the field of the microgrid [28, 29, 30]. However, only the first group of optimal results will be implemented by MPC. More useful information derived by MPC should be explored to maximize the consumption of DG, except only the first one. Besides, compared with medium-time optimization dispatch, energy buffer of users in ER-based ADN can be shared to alleviate the power variation in a short-time scale.

---

*Corresponding author
E-mail address*: mchxia@bjtu.edu.cn (M.C. Xia)

Therefore, an ER with five-layer structure will be proposed, and called grid ER (GER) in this paper. Attention on structure of GER will be paid to achieve peer-to-peer energy sharing among NERs, and the issues encountered by the present ADN will be fully taken into consideration. A GER-based ADN will be developed eventually to achieve peer-to-peer energy sharing. To alleviate the mismatch between the medium-time dispatch and device-level control caused by the DG, the bi-level primary energy dispatching strategy will include two parts: the dispatch strategy based on modified MPC (MMPC) in short-time scale and the distributed tracking strategy combined with FLC in real-time scale. The former will make full use of energy buffer within the entire NERs-based system to minimize the mismatch between the medium-time dispatch and device-level control. The latter will be devoted to sharing variation power within GER-based ADN in a distributed way to better track the MMPC-based strategy. The contributions of this paper are as follows:

1) A hierarchical structure of GER is proposed, and peer-to-peer energy sharing of independent operators with different needs can be conveniently achieved. It has high flexibility on energy management and great expansibility on the structure.

2) Proposed GER can be placed at critical buses of ADN, and a practical scheme of the GER based on five-layer structure is proposed and validated.

3) An MMPC-based dispatch strategy is proposed to make full use of energy buffer within the entire system in short-time scale. More information of optimal results at each dispatch period is further explored to optimize the energy buffer.

4) A distributed tracking strategy is proposed to better manage the SOC of energy buffer device, considering measured power variation in real-time scale. It can make full use of the power variation within the entire GER-based system to better realized the MMPC-based strategy.

The paper is organized as follows. Section 2 presents the proposed structure and function of GER. The primary energy dispatching strategy of GER is presented in section 3. Section 4 specifies the simulation cases and results analysis. Finally, the conclusion is summarized in section 5.

## 2. The hierarchical structure of GER

### 2.1. The structure design of GER

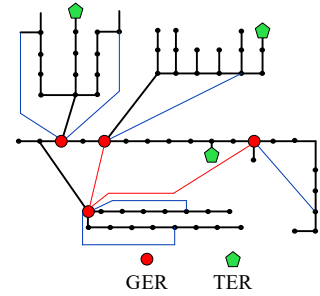

**Fig. 1.** The GER-based ADN integrated with the TER

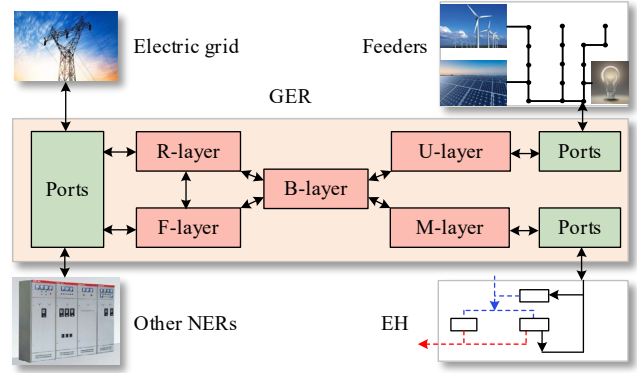

**Fig. 2.** The proposed hierarchical structure of GER

A GER-based ADN with TERs is shown in Fig. 1. Under such application, the NERs are placed at some critical buses and govern a relative larger area than TER. Meanwhile, peer-to-peer energy sharing can be easily achieved between NERs. Closed-loop operation and flexible power flow control of complex ADN will also be achieved. That will help to resolve the issues such as over or under voltage of buses, load imbalance, reverse power flow. Considering the coexistence of AC and DC in future ADN, the GER should be suitable for the AC and DC hybrid systems. In addition, considering the interconnection of other carrier energy such as heat and cold, GER should provide the corresponding interface. Consequently, the design requirements of ADN for GER are: (a) peer-to-peer energy transmission among NERs and autonomous operation within GER can be conveniently and flexibly achieved. (b) GER can be deployed at critical buses to realize the flexible control of power flow within ADN. (c) It is suitable for AC and DC hybrid ADN and multi-carrier energy system. In information Internet, the hierarchical structure of information routers greatly facilitates its development and expansibility, and thus the structure of GER will be scheduled as a multi-layer structure.

Combined with the design of TER, the hierarchical structure of GER proposed in this paper is shown in Fig. 2. The proposed hierarchical structure of GER is mainly catalogued as five layers: (a) Routing layer (R-layer). Energy routing between GER and the external AC or DC subsystem can be derived through R-layer. (b) Forwarding layer (F-layer). When the ancillary services, such as peer-to-peer energy forwarding, is requested from one external GER to another one, F-layer will be deployed to achieve that target if GER is in the transmission route. (c) Buffer layer (B-layer). It can be employed to maintain the real-time energy balance within the entire GER. Compensation of the loss when energy is shared, and suppression of power fluctuation, and sharing energy storage with other GER can also be satisfied by appropriate control of B-layer. Since EV has the ability of charging and discharging, the EV station can be connected at the B-layer to provide auxiliary service. (d) User layer (U-layer). This is employed to connect feeders with PV stations, wind farms, fuel cells, and different kinds of loads. (e) Multi-carrier energy layer (M-layer). Energy exchange between GER and energy hub (EH) can be achieved by M-layer, providing interface for multi-carrier energy sources [26, 30]. The energy conversion equipment can be a boiler and other energy conversion devices.

Compared with the typical ER aiming to the integration of DG and load, the design of proposed GER is more inclined to peer-to-peer energy sharing among NERs. Such a hierarchical structure of GER expands the functions of ER to the aspects on power flow congestion, safety operation constraints of network, energy transaction. The effects the effects of the implementation of the GER on the power distribution reliability can refers to [10], and is omitted.

### 2.2. A feasible system design of the proposed GER
#### 2.2.1 A feasible structure

In this paper, peer-to-peer energy sharing of GER is the research focus of the structural design. There have already been many studies on the accessibility of the U-layer in microgrids and TER, and the M-layer is not the focus of this paper. Thus, both layers are not mentioned in detail, and only one port is deployed for simplicity, respectively.

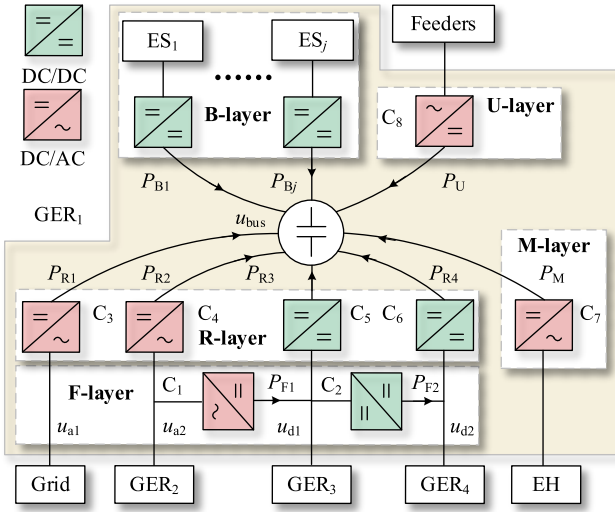

**Fig.3.** The proposed feasible system of GER

A feasible topology of GER and basic configuration are proposed as shown in Fig. 3, including all five layers. $u_{a1}$, $u_{a2}$ are phase-to-phase voltages of corresponding ports of converters $C_3$, $C_4$, respectively. $u_{d1}$ and $u_{d2}$ are the DC voltages of $C_5$ and $C_6$ in R-layer, respectively. $u_{bus}$ is the DC bus voltage. $P_{R1}$, $P_{R2}$, $P_{R3}$, and $P_{R4}$ are the input power of R-layer, respectively. $P_{F1}$, $P_{F2}$ are the forwarding power between the neighbor system in the F-layer respectively. $P_M$, $P_U$ and $P_B$ are the equivalent input powers of M-layer, U-layer and B-layer respectively. Each arrow indicates the reference direction of power. The AC/DC and DC/DC device can be a bidirectional converters, and typical three-phase full bridge inverters and buck-boost converters are employed in the case study in this paper [23, 32].

The features of the feasible topology of proposed GER are as follows: (a) DC link is employed to achieve interconnection and decoupling between different forms of electric energy, which provides convenience for the expansion of ports in each layer. (b) DC and AC ports are provided so that GER is suitable for the coexistence of AC and DC hybrid ADN in the near future. That will facilitate the realization of the energy sharing between itself and the external grid or neighbor NERs. (c) Devices in the F-layer are located between the ports in the R-layer and are relatively independent of the R-layer. That will facilitate relatively independent control in the R-layer and F-layer, as well as efficiency improvement of the F-layer. Energy forwarding in different regions can be implemented between connected AC and AC systems, connected DC and AC systems, and connected

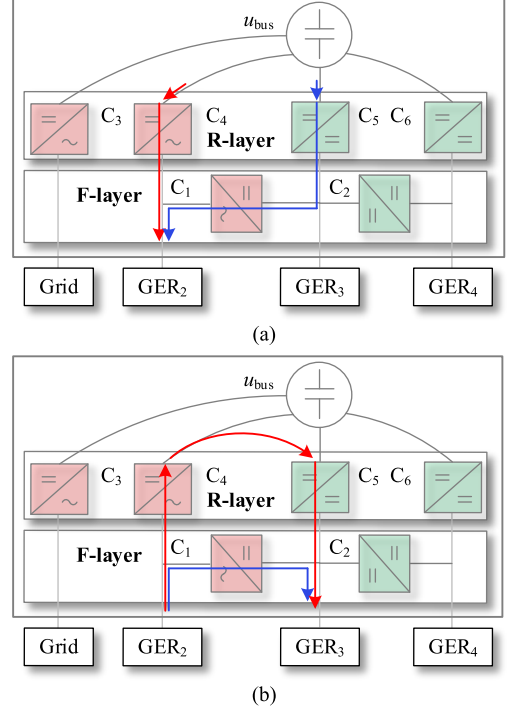

**Fig.4.** The devices are backups for each other. (a) The idle F-layer is backup for the R-layer, (b) The idle R-layer is backup for the F-layer.

DC and DC systems through the F-layer. (d) The R-layer and F-layer are not strictly independent. Energy forwarding between connected AC and AC systems is implemented by corresponding ports (such as $C_3$ and $C_4$) in the R-layer, considering the issue of efficiency. (e) The devices of the R-layer and the F-layer can be backups for each other to minimize the impact of faults on GER. As shown in Fig. 4(a), when energy is routed by $C_4$ and fault occurs in $C_4$, energy routing can be maintained through idle devices $C_1$ and $C_5$. Similarly, as shown in Fig. 4(b), when energy is forwarded by $C_1$ and fault occurs in $C_1$, energy forwarding can be maintained through the idle devices $C_4$ and $C_5$. (f) The B-layer is an indispensable layer in GER, which has strict requirements on reliability and storage capacity. Multiple energy buffer devices in parallel can be employed to facilitate the capacity expansion of B-layer and improve its reliability.

Therefore, the main functions of proposed GER are: (a) energy routing with external systems can be derived through AC and DC ports in the R-layer. (b) Energy forwarding between connected AC and AC systems, connected DC and AC systems, and connected DC and DC systems can be achieved through F-layer. (c) The R-layer and F-layer are backups for each other. (d) Energy buffering ability is equipped at the B-layer. Thus, peer-to-peer-to-peer energy sharing can be conveniently implemented under the cooperation of the R-layer, F-layer and B-layer.

*2.2.2 A feasible systematic control configuration*

According to the GER-based ADN and hierarchical structure of GER shown in Fig. 1 and Fig. 3 respectively, a feasible systematic control scheme should be proposed to achieve stable operation of the GER-based system. The stability of $u_{bus}$ within GER should be maintained as shown in (1).

$$C_{dc} du_{dc} / dt = \sum P_{Ri} + \sum P_{Bj} + P_U + P_M \qquad (1)$$

$C_{dc}$ is the equivalent capacitance of the DC link. When the right side of (1) is maintained to 0, $u_{bus}$ will be stable and the GER will operate stably. In this paper, the B-layer is deployed to control $u_{bus}$, and thus the power fluctuation caused by other layers of the GER is suppressed by the B-layer first.

The typical control strategies of AC/DC converter include $V/f$ control, $PQ$ control and $u_{dc}Q$ control [7, 23]. Reactive power employed to support AC voltage and exchange of active power between AC and DC sides can together be achieved by $PQ$ control. $V/f$ control can be utilized to regulate the frequency and voltage of the AC system. Typical control strategies of DC/DC converter include current control, $P$ control and $U$ control [23, 32]. Power exchange between two sides of the converter can be achieved by $P$ control, and voltage can be maintained at a certain configured value by $U$ control. In order to simplify the analysis, assume that the voltage or frequency of the corresponding port of R-layer is formed by the external system. Therefore, $PQ$ control is adopted by $C_1$, $C_3$ and $C_4$, while $P$ control is employed by $C_2$, $C_5$ and $C_6$. $V/f$ control is deployed by DER$_2$, and $U$ control is utilized by DER$_3$ and DER$_4$ at corresponding port, respectively. For the DER$_1$, $V/f$ control or $U$ control also can be configured at corresponding ports if neighboring NERs adopt $PQ$ control or $P$ control. The difference is which side implements the instructions of the power exchange.

It is worth noting that the feasible system of proposed GER can be easily modified by adding or reducing the corresponding devices at each layer according to the application. For example, when GER is applied to pure DC DN, the corresponding AC port can be omitted. When more storage capacity is required in the B-layer, more energy buffer devices can be installed. Therefore, the structure of the feasible system has great flexibility and expansibility, and thus can be employed to achieve peer-to-peer energy sharing.

## 3. Primary energy dispatching strategy of the proposed GER

The primary energy dispatching strategy is proposed only for the energy management within the GER instead of the GER-based ADN. Thus, the power flow constraints is omitted for simplicity. Assume that the power variation imposed by DGs and loads at the U-layer is handled by the energy buffer of GER in short-time and real-time scales, and the controllable devices at the U-layer just implement the medium-time dispatch. The proposed strategy will include two parts: the MMPC-based dispatch strategy and the distributed tracking strategy in real-time scale. Both parts interact with each other to alleviate the mismatch between the medium-time dispatch and device-level control caused by the forecast error of DG.

*3.1. The MMPC-based dispatch strategy*

*3.1.1 Energy management of energy buffer*

Since the energy buffer of the B-layer is finite considering the cost, it is rational that the excessive power variation is consumed directly by the energy buffer of neighbor NERs. Thus, some GER may own relatively larger energy buffer due to the variation of DG, and are tolerant of a relative large range of power variation from the adjacent GER. Some NERs may have less energy capacity, and thus the energy sharing should be implemented strictly according to medium-time dispatch. Therefore, the MMPC-based dispatch strategy is proposed to alleviate the mismatch between the medium-time dispatch and device-level control caused by the DG in short-time scale through energy buffer of the B-layer within the entire system.

For the dispatching power $P_{Ri}^{ref}$ of port $i$ at R-layer derived from medium-time dispatch, certain variation range $[-\alpha_i, \alpha_i]$ is expected when energy is shared, taking into account different needs of users and maximum consumption of DG. The bigger variation range means larger energy capacity. For simplicity, the power variation from U-layer only includes loads, PV, and wind turbine (WT), i.e. $\Delta P_{Load}$, $\Delta P_{PV}$, and $\Delta P_{WT}$, respectively. According to proposed systematic control strategy, such variation power will be suppressed by the B-layer first. In order to better implement MMPC-based strategy within the entire GER-based ADN, the GER is catalogued as the master GER and slave one. The shared energy is implemented by the master GER, and the slave GER just receives or transmits the energy passively according to the control of the master GER. Thus, the MPC-based dispatching strategy is shown in (2)~(10).

$$\min \sum_{k=t_0}^{t_0+T-1} \sum_{i=1}^{N_s} \omega_i J_i^2(k) + \gamma \sum_{k=t_0}^{t_0+T-1} (\sum_{i=1}^{N_s} P_{Ri}(k) + P_B(k) + P_U(k))^2 \qquad (2)$$

$$J_i = \begin{cases} P_{Ri} - (1+\alpha_i)P_{Ri}^{ref}, \text{if } P_{Ri}^{ref} > 0 \text{ and } P_{Ri} > (1+\alpha_i)P_{Ri}^{ref} \\ P_{Ri} - (1-\alpha_i)P_{Ri}^{ref}, \text{if } P_{Ri}^{ref} > 0 \text{ and } P_{Ri} < (1-\alpha_i)P_{Ri}^{ref} \\ P_{Ri} - (1-\alpha_i)P_{Ri}^{ref}, \text{if } P_{Ri}^{ref} < 0 \text{ and } P_{Ri} > (1-\alpha_i)P_{Ri}^{ref} \\ P_{Ri} - (1+\alpha_i)P_{Ri}^{ref}, \text{if } P_{Ri}^{ref} < 0 \text{ and } P_{Ri} < (1+\alpha_i)P_{Ri}^{ref} \\ 0, \text{ otherwise} \end{cases} \qquad (3)$$

$$\sum_{i=1}^{N_s} P_{Ri} + P_B + P_U = 0 \qquad (4)$$

$$P_U = \sum \Delta P_{PV} + \sum \Delta P_{WT} - \Delta P_{Load} + \sum_{i=1}^{N_s} P_{Ri}^{ref} \qquad (5)$$

$$E_B(k) = E_B(k-1) - \Delta T \cdot P_{B\_dis}(k) / \eta_{dis} - \Delta T \cdot P_{B\_ch}(k) \times \eta_{ch} \qquad (6)$$

$$\begin{cases} P_{B\_dis}(k) = P_B(k), \text{ if } P_B(k) > 0 \\ P_{B\_ch}(k) = P_B(k), \text{ else} \end{cases} \qquad (7)$$

$$\begin{cases} E_{B\_min} \leq E_B(k) \leq E_{B\_max} \\ -P_{BN} \leq P_B \leq P_{BN} \end{cases} \qquad (8)$$

$$-P_{RNi} \leq P_{Ri} \leq P_{RNi} \qquad (9)$$

$$E_{B\_init} = E(t_0 - 1) \qquad (10)$$

$N_s$ is the number of ports in the R-layer. $\Delta T$ is the time step. $T$ is the total time period at each optimization process. Constraint (3) is set to generate power with different kind of variation permitted by customers to suppress power variation in a short-time scale. For the master GER, weight value $\omega_i$ is set larger to give priority to the ports of the R-layer with strict power variation, and the corresponding variation range should be satisfied first. In this paper, $\omega_i$ is set to $1/\alpha_i$ ($\alpha_i > 0$). Meanwhile, $\gamma$ is set to 0 to decrease the degree of state space. For the slave GER, $P_{Ri}$ is uncontrollable. In such a case, $\omega_i$ is set to 0 and $\gamma$ is set to 1. It is worth noting that other strategies such as load control $\Delta P_{Load}$ may be required to meet constraints (4) for slave GER. Constraints (6)~(8) are set to meet the requirements of life extension. Constraint (10) is the state of B-layer and required for MPC. The proposed MPC-based dispatch strategy of the GER is constraint and nonlinear model, and thus is solved by the genetic algorithm (GA) in this paper [33].

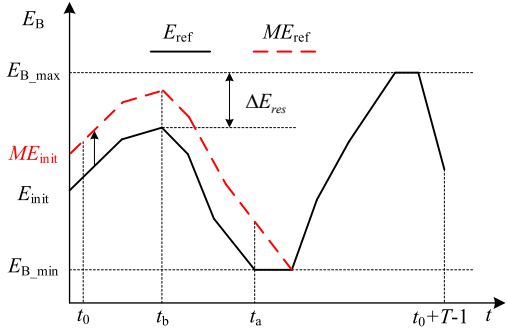

**Fig.5.** Shifting the instruction of energy buffer after the optimization

*3.1.2 Optimized energy buffer instruction*

For typical MPC, only the instructions at the first time interval will be implemented. However, according to the instructions in the rest time intervals, more optimal dispatch may be derived with better $E_{init}$ of the B-layer at each dispatch process, as shown in Fig. 5. When $E_{ref}$ of the B-layer reaches its minimum limit $E_{B\_min}$ at $t_a$, and there is enough capacity $\Delta E_{res}$ at its peak time $t_b$ ($t_b \in [t_0, t_a]$), a upward shift of the $E_{ref}$ can improve the optimization results at $t_a$ if power variation cannot be suppressed totally. The improvement is constrained within $[t_0, t_a]$ and should not affect the rest of optimization results. The best adjustment of $E_{ref}$ should be the red dashed line, and the changed amount of energy is $\Delta E_{sh}$. Thus, the MPC is modified and shown in (11)~(14).

$$ME_{ref}(k) = \begin{cases} E_{ref}(k) + \Delta E_{sh}, & if\ k \in [t_0, t_a] \\ E_{ref}(k), & else \end{cases} \quad (11)$$

$$\Delta E_{sh} = \begin{cases} \min\left\{\dfrac{1}{\eta_{dis}} \cdot \Delta T \sum_{k=t_0}^{k=t_a} \sum_{i=1}^{N_s} J_i(k), \Delta E_{res}\right\}, \\ \quad if\ E_{ref}(t_a) = E_{B\_min} \\ \max\left\{\eta_{ch} \cdot \Delta T \sum_{k=t_0}^{k=t_a} \sum_{i=1}^{N_s} J_i(k), \Delta E_{res}\right\}, \\ \quad if\ E_{ref}(t_a) = E_{B\_max} \\ 0, otherwise \end{cases} \quad (12)$$

$$\Delta E_{res} = \begin{cases} E_{B\_max} - E_{ref}(t_b), \\ if\ E_{ref}(t_a) = E_{B\_min}\ and\ E_{ref}(t_b) = \max_{k \in [t_0, t_a]}(E_{ref}(k)) \\ E_{B\_min,} - E_{ref}(t_a), \\ if\ E_{ref}(t_a) = E_{B\_max}\ and\ E_{ref}(t_b) = \min_{k \in [t_0, t_a]}(E_{ref}(k)) \end{cases} \quad (13)$$

$$t_a = \min\left\{\arg_k E_{ref}(k) = E_{B\_min}, \arg_k E_{ref}(k) = E_{B\_max}\right\} \quad (14)$$

Therefore, the entire MMPC-based dispatching strategy of the GER is shown in (2)~(14).

To evaluate the performance of the MMPC-based dispatching strategy of the GER, the uncomfortable index $UC_i$ of port $i$ at R-layer is shown in (15).

$$UC_i = \sum_{k=t_0}^{t_0+T-1} (J_i(k)/P_{Ri}^{ref}(k))^2 \quad (15)$$

It is worth noting that the optimal results may be different from those calculated by (11) if the MMPC-based dispatching strategy is resolved with $\Delta E_{sh}$. That doesn't matter because one of the main targets of MMPC is to provide better $E_{ref}$ for the bottom tracking strategy. At the next optimal dispatch period in the short-time scale, $E_{init}$ will be updated and utilized by the MMPC-based dispatching strategy. Besides, only the reference energy of B-layer is modified, and other instructions such as $P_{Ri}$ should be implemented directly.

*3.2. The distributed tracking strategy*
*3.2.1 Energy buffer instruction tracking*

The power variations in real-time scale are relatively complementary. For the stochastic features of different power variations, those fit certain kinds of distribution functions [34, 35, 36]. In this paper, the probability density function (PDF) of the load and WT are selected as Gaussian distribution function to represent the forecast error in real-time scale, and that of PV power generation is Beta distribution function. The target of the distributed tracking strategy is to make the B-layer track its modified reference $ME_{ref}$ through sharing variation power within the GER-based ADN in real-time scale. Less calculation time should be one of the main features for a distributed iteration process to achieve rapid energy exchange in the real-time scale. Thus, FLC is employed to achieve multiple targets.

There are two input variables and one output variable for the proposed FLC-based strategy. The defuzzification method is the typical centroid method [22]. The first input is deviation $\Delta E_{ref}$ between real-time value of $E_B$ and $ME_{ref}$. When $\Delta E_{ref}$ is positive, the energy shortage occurs. Otherwise, the surplus energy occurs. $\Delta E_{ref}$ is shown in (16).

$$\Delta E_{ref} = ME_{ref} - E_B \quad (16)$$

The second input variable is the emergency factor $m_a$. When $m_a$ is close to 0, the operation time is very close to the modified time $t_a$ and $\Delta E_{ref}$ should be relatively small to better realize the target of proposed MMPC-based strategy. $m_a$ is shown in (17).

$$m_a = (t_a - t)/T \quad (17)$$

The output variable is the expected exchange power $NP_c$. In GER-based ADN, each GER will generate its $NP_c$. If $NP_c$ is positive, energy should be transmitted to other NERs with negative $NP_c$. Otherwise, energy should be inputted to itself from other NERs with positive $NP_c$. Table I shows the quantitative results of inputs and output. $E_{bd}$ and $P_{cN}$ are the corresponding rate values of $\Delta E_{ref}$ and $NP_c$, respectively.

**Table 1**
Quantitative results of inputs and output

| Variable | Basic Domain | Fuzzy Domain | Fuzzy Subset |
|---|---|---|---|
| $\Delta E_{ref}$ | $[-E_{bd}, E_{bd}]$ | [-2 -1 0 1 2] | [NB NS ZO PS PB] |
| $m_a$ | [0 1] | [0 1 2 3 4] | [C SC M SF F] |
| $NP_c$ | $[-P_{cN}, P_{cN}]$ | [-2 -1 0 1 2] | [NB NS ZO PS PB] |

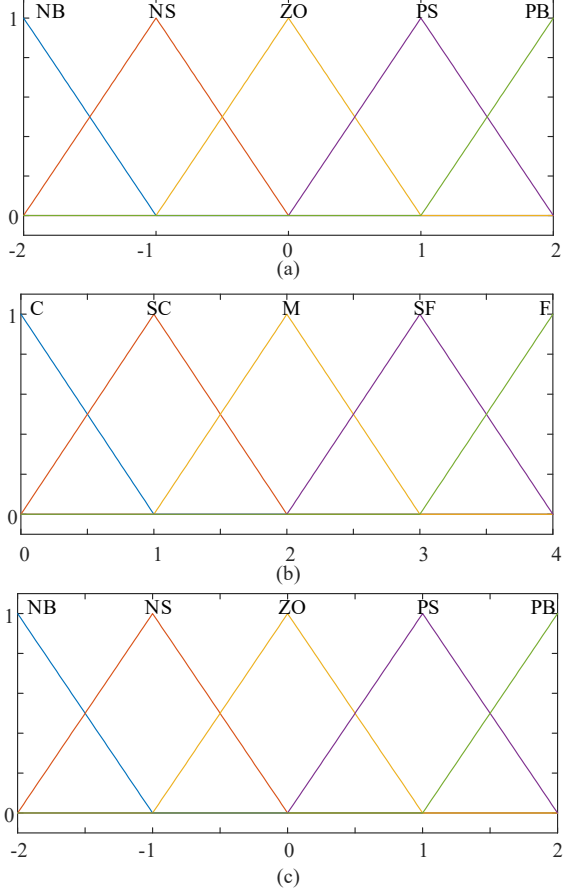

**Fig. 6.** Membership functions of input variables and output variable. (a) $\Delta E_{ref}$, (b) $m_a$, (c) $NP_c$.

For simplicity, five grades are employed to describe $m_a$: C (close), SC (slightly close), M (medium), SF (slightly far), F (far). For $\Delta E_{ref}$ and $NP_c$, they are also denoted by five grades: NB (negative big), NS (negative small), ZO (zero), PS (positive small), PB (positive big). The membership functions of the input and output variables are shown in Fig. 6.

Table II shows the fuzzy control rules, which are designed according to the following basic criterion:

1) When $m_a$ is very small, it means that the energy stored in the B-layer is not enough at the very close time interval. Thus, expected energy to be compensated by other NERs should be large. Otherwise, expected energy should be relatively small.

2) When $\Delta E_{ref}$ is negative, it means that energy stored in the B-layer is more than that expected. Thus, the B-layer should be discharged. Otherwise, the B-layer should be charged to track its reference $ME_{ref}$.

It is worth to be noted that for the power flow constraint of network, only the port current limit is taken into consideration for simplicity in this paper. If the $P_{RN}$ is violated by $NP_c$ in the R-layer, the final power of R-layer should be limited by master GER. Besides, $NP_c$ should be modified to $P_c$ considering the $P_{BN}$ in the B-layer as shown in (18).

$$P_c = \begin{cases} \max(-P_{BN} - P_B, \min(NP_c, P_{BN} - P_B)), & \text{if } P_B > 0 \\ \min(P_{BN} - P_B, \max(NP_c, -P_{BN} - P_B)), & \text{else} \end{cases} \quad (18)$$

**Table 2**
Fuzzy logic rule base

| $NP_c$ | | $\Delta E_{ref}$ | | | | |
|---|---|---|---|---|---|---|
| | | NB | NS | ZO | PS | PB |
| $m_a$ | C | PB | PS | ZO | NS | NB |
| | SC | PB | PS | ZO | NS | NB |
| | M | PB | PS | ZO | NS | NB |
| | SF | PS | ZO | ZO | ZO | NS |
| | F | PS | ZO | ZO | ZO | NS |

*3.2.1 Distributed energy sharing process*

The compensated energy $P_c$ of all NERs can be derived by FLC, but actual compensated energy should be carried out only when the expected compensation is complementary. Thus, a fast compensation strategy (FCS) to achieve the compensation conveniently and rapidly should be put forward, and $P_c$ will be modified to $MP_c$.

Considering an undirected graph $G(V, E)$ formed by a GER-based ADN with $s$ NERs, the vertex set $V$ denotes all NERs and edge set $E$ denotes the electrical connections of adjacent NERs. The adjacent matrix of $G$ is $A = (a_{i,j})_{s \times s}$. If there exists an electrical connection between $DER_i$ and $DER_j$, then $a_{i,j} = a_{j,i} = 1$. Otherwise, those value should be 0. $a_{i,i}$ is 0 all the time. The compensation vector is $P_c = [P_{c1}, P_{c2}, \ldots, P_{cs}]$ and divided into $P_{nn}$ with non-positive elements and $P_{pp}$ with nonnegative elements. The relationship is shown in (19).

$$P_{nn} + P_{pp} = P_c \quad (19)$$

The detailed processes of FCS are as follows:
(a) $P_c$ of NERs are shared with each other.
(b) $DER_i$ with nonnegative compensation shares its energy to adjacent $DER_j$ with negative compensation. Then GER $j$ can obtain expected compensation $B = (b_{i,j})_{s \times s}$ shown in (20).

$$b_{i,j} = \begin{cases} 0, & \text{if } \sum_{g=1}^{s}(a_{j,g} \times P_{nng}) = 0 \\ P_{ppj} \times a_{j,i} \times P_{nni} / \sum_{g=1}^{s}(a_{j,g} \times P_{nng}), & \text{else} \end{cases} \quad (20)$$

(c) When GER $j$ with negative compensation obtains all expected compensations from adjacent NERs with positive compensation, $B$ is modified into $C = (c_{i,j})_{s \times s}$ shown in (21).

$$c_{j,i} = \begin{cases} b_{j,i} \times P_{nnj} / \sum_{h=1}^{s} b_{j,h}, & \text{if } \sum_{h=1}^{s} b_{j,h} > |P_{nnj}| \text{ and } \sum_{h=1}^{s} b_{j,h} \neq 0 \\ b_{j,i}, & \text{else} \end{cases} \quad (21)$$

Finally, the FCS is shown in (19)~(21). Each column of $C_{s \times s}$ includes modified power of each port for the R-layer. If the value is negative, the instruction should be increased. Otherwise, the instruction should be decreased. For the B-layer, the compensating power $MP_c$ is shown in (22).

$$MP_c(i) = \sum_{j=1}^{j=s} c(i,j) \tag{22}$$

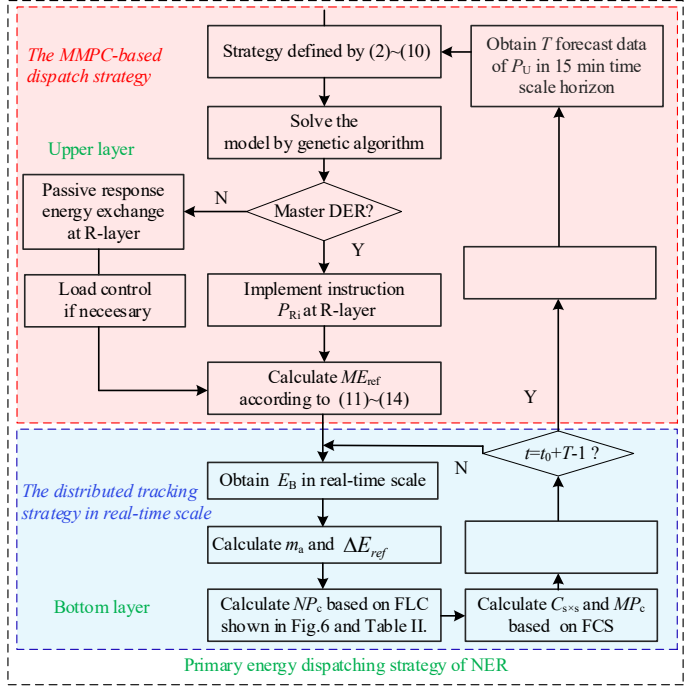

**Fig. 7.** The entire process of proposed primary energy dispatching strategy

Therefore, the proposed distributed tracking strategy in the real-time scale is totally proposed. The entire process of proposed primary energy dispatching strategy of proposed GER in GER-based network is shown in Fig. 7.

## 4. Case study

To verify the proposed hierarchical structure and primary energy dispatching strategy of GER, three scenarios are designed: *1) Verification of proposed MMPC-based dispatch strategy*. Verify that different needs of energy exchange can be satisfied, and MMPC-based strategy can achieve better optimization results. *2) Verification of proposed distributed tracking strategy*. Verify that $ME_{ref}$ and variation of DG can be better tackled with proposed distributed strategy. *3) Verification of proposed hierarchical structure of GER*. Verify that peer-to-peer energy sharing can be achieved by the proposed system. Meanwhile, proposed primary energy dispatching strategy can be implemented at the proposed GER conveniently.

### 4.1. Verification of proposed MMPC-based dispatch strategy

**Table 3**
Main simulation parameters in case 4.1

| $P_{BN}$(kW) | $P_{RN2}$(kW) | $P_{RN3}$(kW) | $E_{B\_min}$(kWh) |
|---|---|---|---|
| 80 | 160 | 160 | 10 |
| $\alpha_2$ | $\alpha_3$ | $E_{B\_max}$(kWh) | $\eta_{ch}, \eta_{dis}$ |
| 0.2 | 0.05 | 40 | 0.9 |
| | Sub-case 1 | Sub-case 2 | Sub-case 3 |
| $E_{init}$ (kWh) | 37 | 12 | 23.42 |

In this scenario, three subcases will be verified: 1) subcase 1. Different needs can be satisfied. 2) Subcase 2. Energy sharing with small power variation range $\alpha_i$ can be satisfied in priority.

3) Subcase 3. Optimization results can be improved by MMPC-based strategy. The time interval for each data is 15 minutes. The control of the slave GER is easy to be implemented, and the verifications focus on the master GER.

Assume that port 2 and 3 of R-layer in $DER_1$ are utilized to share energy with $DER_2$ and $DER_3$. The main simulation parameters are shown in Table III.

#### 4.1.1 Sub-case 1

For simplicity, only 24 data points are utilized to verify the effectiveness. The simulation results are shown in Fig. 8. $P_{R2-b1}$ and $P_{R2-b2}$ are expected variation bounds of port 2 of R-layer. $P_{R3-b1}$ and $P_{R3-b2}$ are that of port 3. The uncomfortable index $UC$ of port 2 and port 3 are 0. $\Delta E_{sh}$ is 0. Thus, both of the expected energy are satisfied completely. Therefore, different needs of NERs are satisfied by proposed MMPC-based dispatch strategy under sufficient energy capacity of the B-layer.

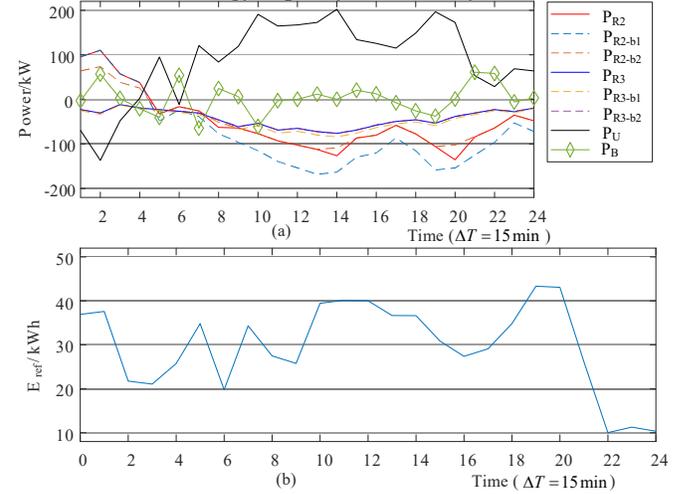

**Fig. 8.** Simulation results of sub-case 1

It is worth noting that once optimization process is enough to verify proposed strategy. Thus, only once optimization process is implemented unlike that for verification of MPC itself under medium-time dispatch.

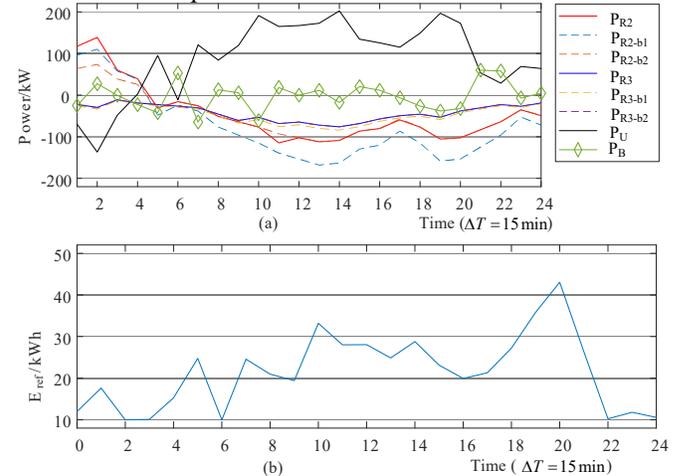

**Fig. 9.** Simulation results of sub-case 2

#### 4.1.2 Sub-case 2:

Compared with sub-case 1, the $E_{init}$ of the B-layer is reduced to 12 kWh. The simulation results are shown in Fig. 9. $P_{R2-b1}$ is clearly violated by $P_{R2}$ at time interval [1 2]. $t_a=2$, and $t_b=1$.

$\Delta E_{sh}$ is 11.42 kWh. Thus, insufficient energy capacity of B-layer occurs at the second point. The calculated uncomfortable index $UC_2$ and $UC_3$ are 0.1694 and 0.0013 respectively. Thus, the simulation results verify that energy sharing with small power variation range $\alpha_i$ can be satisfied in priority under insufficient capacity of the B-layer.

*4.1.3 Sub-case 3*

According to the compensation capacity derived by subcase 2, the $E_{init}$ of B-layer is added with 11.42, and thus its value becomes 23.42 in subcase 3. The simulation results are shown in Fig. 10. The curve of $P_{R2}$ is clearly improved at time interval [1 2]. $t_a=6$, and $t_b=1$. $\Delta E_{sh}$ is 2.627 kWh. That means a slightly insufficient energy capacity of the B-layer at the sixth point. The calculated uncomfortable index $UC_2$ and $UC_3$ are 0.0063 and 0.0001 respectively. Thus, the simulation results verify that optimization results can be improved by proposed MMPC-based strategy.

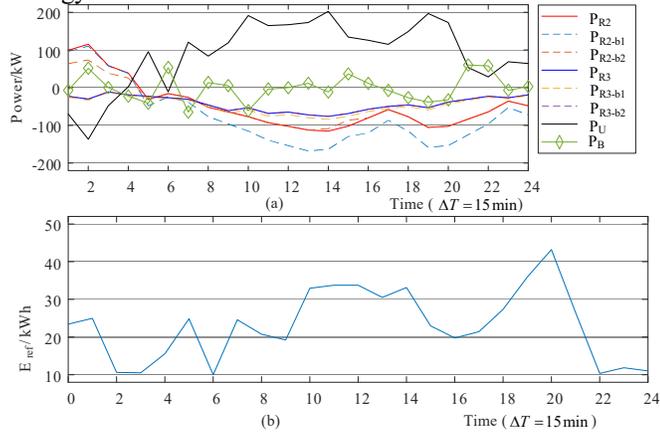

**Fig. 10.** Simulation results of sub-case 3

Therefore, the above simulation results verify that different needs of NERs can be better satisfied by the proposed MMPC-based dispatch strategy when the initial SOC is improved during specified time slots.

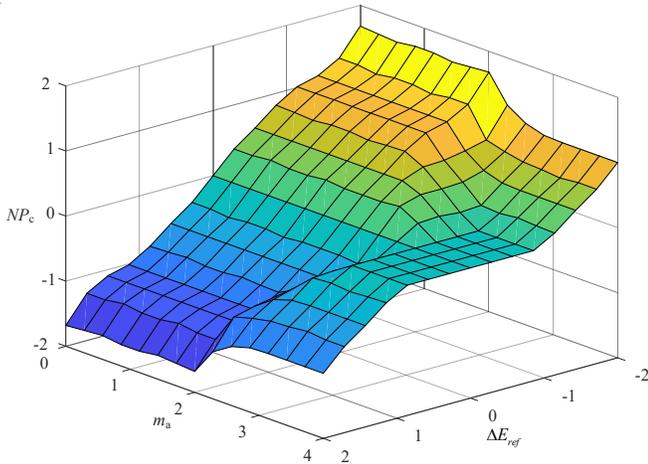

**Fig. 11.** The designed control surface of FLC

*4.2. Verification of proposed distributed tracking strategy*

The designed control surface of FLC is shown in Fig. 11. The compensation $NP_c$ is well-designed to cooperate with the proposed MMPC-based dispatch strategy.

In order to verify the proposed FLC-based tracking strategy, a 10-GER-based ADN is simulated. The topology is shown is Fig. 12. $E_{bd}$ is set as 30kWh, and $P_{cN}$ is 80 kW. For simplicity, PB is generated by Gaussian random distribution and the port rate current $P_{BN}$ is 80kW. Each time interval for proposed FLC-based tracking strategy is 3 minutes. In order to better verify the proposed strategy, 100 intervals are simulated. That will not affect the validity of the simulation results.

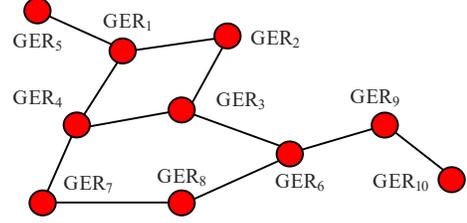

**Fig. 12.** The topology of a 10-GER-based ADN

The simulation results of the proposed FLC-based tracking strategy are shown in Fig. 13. dP1~dP10 are real-time power variations of DER$_1$~DER$_{10}$ respectively. Sh1~Sh10 are the total compensation powers of DER$_1$~DER$_{10}$ respectively. Considering the power variation and the capacity reference, dE1~dE10 are the calculated deviation of buffer capacity without the proposed FLC-based tracking strategy. Considering the power variation and the capacity reference, ddE1~ddE10 are the deviation of buffer capacity with proposed FLC-based tracking strategy. mdE1~mdE10 are the first inputs ($\Delta E_{ref}$) of proposed the FLC-based tracking strategy in DER$_1$~DER$_{10}$ respectively. ma1~ma10 are the second inputs ($NP_c$) of the proposed FLC-based tracking strategy. In Fig. 13(c), the deviations of buffer capacity vary according to only their power variations. By employing the proposed FLC-based tracking strategy, variation power is shared within GER-based ADN according to their own states as shown in Fig. 13(b), Fig. 13(d) and Fig. 13(f). Thus, the deviations range of buffer capacity in B-layer are alleviated and tend to be 0 as shown in Fig. 13(d). It can be concluded that the adverse impact of variation power in real-time scale can be alleviated and be utilized to improve the stored energy in B-layer for MMPC-based dispatch strategy.

Therefore, the simulation results verify that the proposed MMPC-based dispatch strategy can be implemented better by the proposed distributed tracking strategy in real-time scale. Thus, the effectiveness is validated.

It is worth noting that the more complementary of the variation power in different NERs is, the better the results of will be derived. For simplicity, only the port current limit of the GER is taken into consideration for power flow constraint.

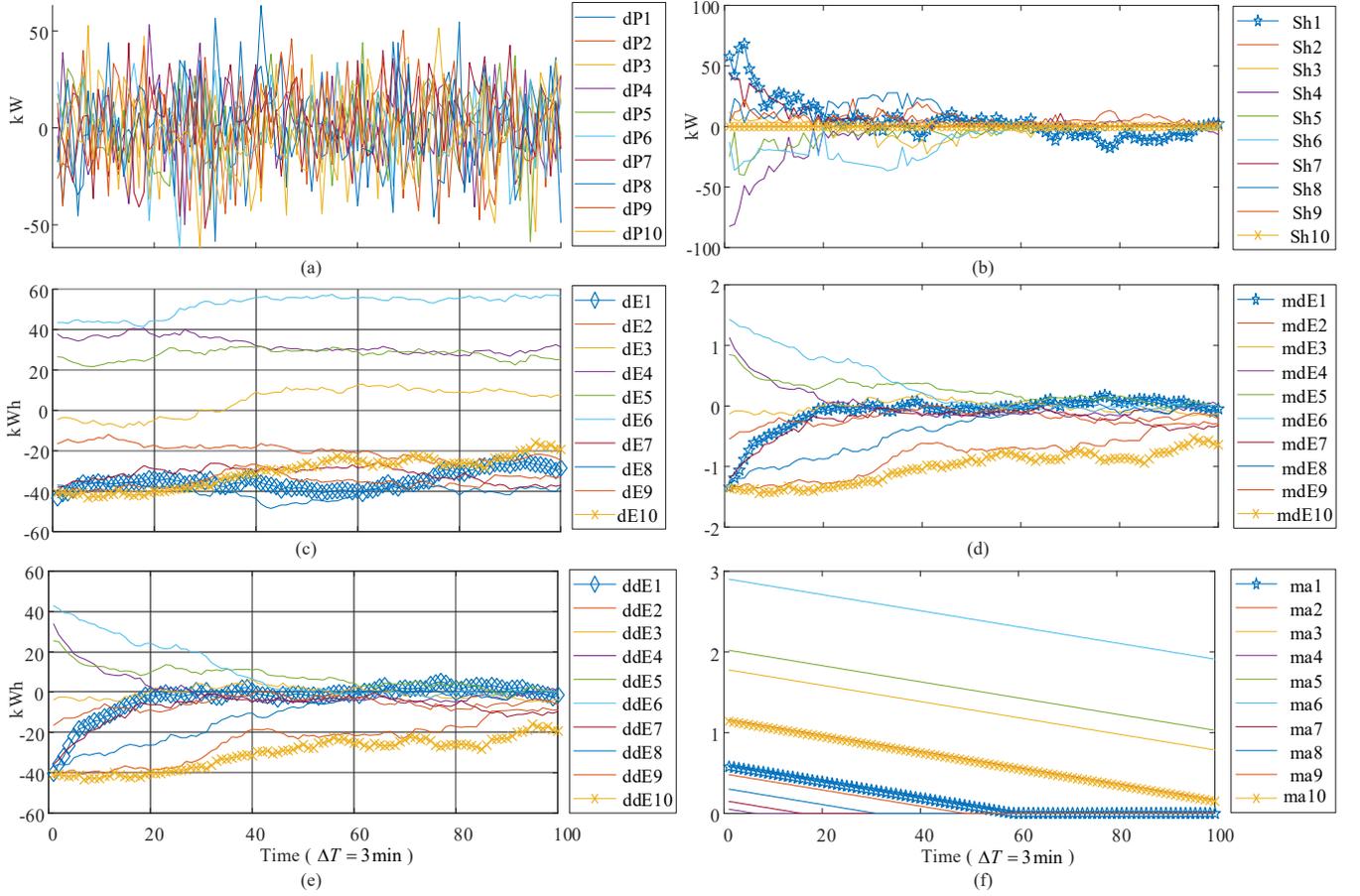

**Fig. 13.** The simulation results of proposed FLC-based tracking strategy. (a) The variation power of each GER in the real-time scale, (b) The values of $MP_c$, (c) The values of $\Delta E_{ref}$ without any energy sharing control, (d) The input values of $\Delta E_{ref}$ in FLC, (e) The values of $\Delta E_{ref}$ with the proposed distributed tracking strategy, (f) The values of $m_a$.

## 4.3. Verification of proposed hierarchical structure of GER.

**Table 4**
Main simulation parameters in case 4.3

|  | $u_{bus}$(V) | $u_{d1}$(V) | $u_{d2}$(V) |
|---|---|---|---|
| Rate voltage | 600 | 700 | 850 |
|  | $u_{a1}$(V) | $u_{a2}$(V) | |
| Rate voltage | 320 | 320 | |

There are two subcases in this scenario: *1) sub-case 1*. Verify that peer-to-peer energy sharing can be achieved by the proposed GER based on hierarchical structure. *2) Sub-case 2*. Proposed primary energy dispatching strategy can be conveniently implemented by proposed GER. The proposed GER shown in Fig. 3 is built and simulated in Matlab/Simulink. Only one port is deployed for the B-layer for simplicity. The main simulation parameters are shown in Table IV. Typical parameters of converters such as inductor filter and capacitor filter are employed and omitted [23, 32]. The circulating current between inverter $C_1$ and $C_4$ is solved by the isolated transformer.

*4.3.1 Sub-case 1*

The simulation results are shown in Fig. 14. In Fig. 14(a), the basic process is that the energy shortage at the U-layer occurs at 0.5s and is expected to be terminated at 1.6s. Because of the forecast error, the energy shortage is terminated at 1.7s. Then the energy surplus is expected to occur at 1.8s, but occurs at 1.9s because of power variation. The surplus energy is terminated at 2.9s. In the entire process, energy sharing is achieved stably and smoothly by connected NERs of the R-layer. The imbalanced power is suppressed by its B-layer in real-time scale. Thus, the simulation results verify that energy exchange among NERs is stable and independent of each other at the R-layer.

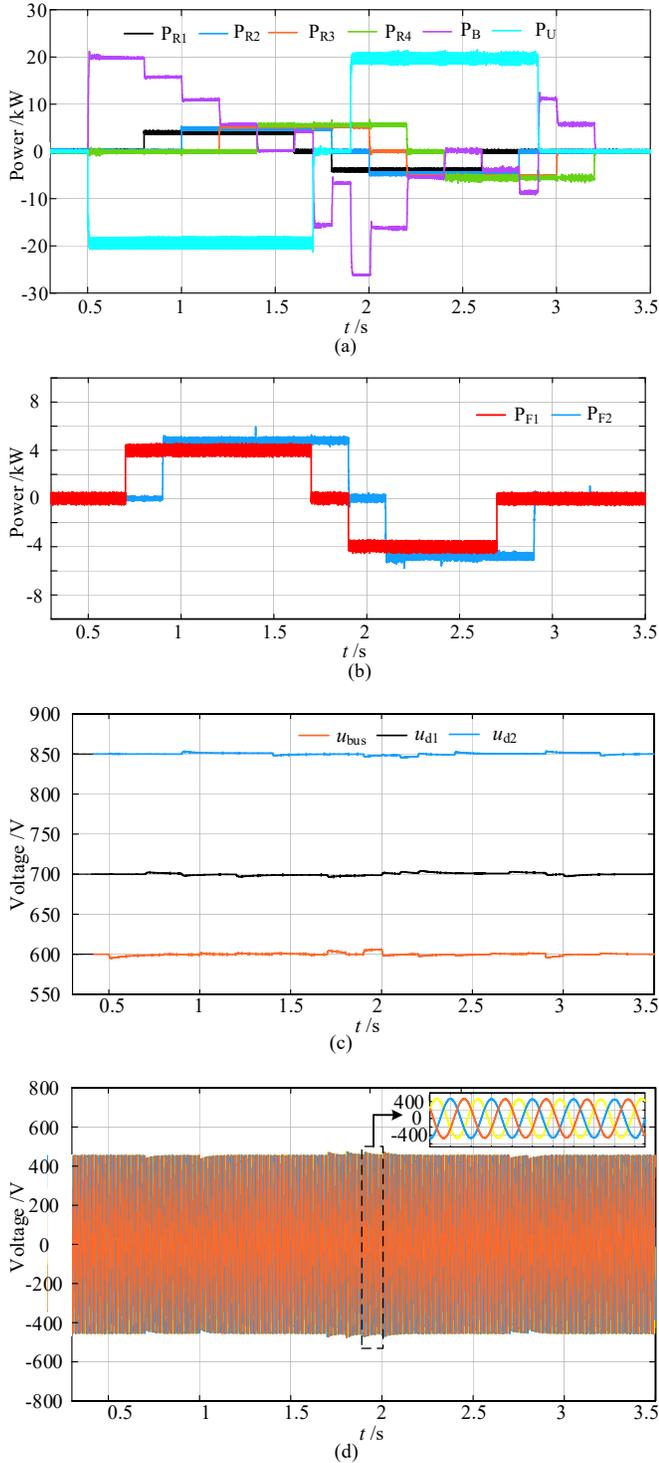

**Fig. 14.** The simulation results of the proposed hierarchical GER. (a) The simulation results of the R-layer, B-layer and U-layer, (b) The simulation results of the F-layer, (c) The simulation results of all DC voltages, (d) The simulation results of $u_{a2}$.

In Fig. 14(b), the basic process is that energy forwarding is implemented by converter $C_1$ and $C_2$ at the F-layer. the simulation results are stable and independent, which verifies the function of F-layer. The simulation results shown in Fig. 14(a) and Fig. 14(b) are simulated together, which verify that the control of R-layer and F-layer are relatively independent.

In Fig. 14(c) and Fig. 14(d), the voltages of corresponding ports are stable and maintained their rate values. Thus, the simulation results verify that peer-to-peer energy sharing can be conveniently achieved by the proposed GER with hierarchical structure. Such structure has high flexibility in energy sharing and power flow control.

*4.3.2 Sub-case 2*

As shown in Fig. 15, the process is that there is an energy surplus at the U-layer. The surplus power is shared by port 1 of the R-layer. At 1.2s, a power variation of 2kW occurs at the U-layer, and energy shortage is confirmed by $DER_2$. Thus, the variation power is transmitted to $DER_2$ during 1.4s and 1.6s. Similarly, the variation power is obtained from $DER_3$ with surplus energy. Thus, the simulation results verify that the proposed primary energy dispatching strategy can be conveniently implemented by the proposed GER.

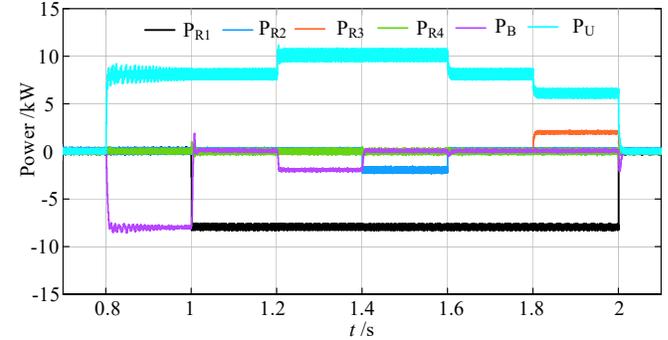

**Fig. 15.** The simulation results of sharing variation power

Therefore, the simulation results verify that peer-to-peer energy sharing can be conveniently achieved by proposed hierarchical structure of the GER, and the proposed primary energy dispatch can be implemented at the proposed GER.

## 5. Conclusion

The implementation of EI requires a flexible ADN to achieve peer-to-peer energy sharing. Therefore, a hierarchical structure and primary energy dispatching strategy of GER are proposed. The proposed primary energy dispatching strategy includes two parts: the MMPC-based strategy and the distributed tracking strategy. The former is proposed to better share the energy buffer among NERs and implement medium-time dispatch. The latter is proposed to better alleviate adverse impact of power variation and to track the MMPC-based strategy. The conclusions are as follows:

1) The peer-to-peer energy sharing of NERs with different needs are conveniently satisfied by the proposed GER with hierarchical structure. The proposed GER has high flexibility on energy management and great expansibility on the structure.

2) A practical scheme of the GER based on five-layer structure is proposed and validated for the proposed primary energy dispatch strategy.

3) The mismatch between the medium-time dispatch and device-level control caused by the forecast error of renewable energy can be alleviated by the proposed primary dispatch strategy. Energy buffer of neighbor NERs in GER-based ADN can be shared and optimized.

4) Relatively complementary power variation within GER-based ADN in the real-time scale is better suppressed and utilized by proposed distributed tracking strategy, and the MMPC-based dispatch strategy is better tracked. Adverse impact imposed by power variation is alleviated.

**CRediT authorship contribution statement**

**Meifu Chen**: Data curation, Methodology, Software, Validation, Writing- Original draft preparation. **Mingchao Xia**: Conceptualization, Methodology, Investigation. **Qifang Chen**: Investigation, Validation.

**Declaration of Competing Interest**

The authors declare that they have no known competing financial interests or personal relationships that could have appeared to influence the work reported in this paper.

**Acknowledgments**

This work was supported in part by National Natural Science Foundation of China (No. 51677003).